\documentclass[english]{achemso}
\usepackage[T1]{fontenc}
\usepackage[latin9]{inputenc}
\usepackage{units}
\usepackage{amsmath}
\usepackage{amssymb}
\usepackage{graphicx}

\makeatletter

\title{StochasticGW-GPU: rapid quasi-particle energies for molecules beyond
10000 atoms}
\author{Phillip S. Thomas, Minh Nguyen, Dimitri Bazile, Tucker Allen, Barry
Y. Li, Wenfei Li, Mauro Del Ben, Jack Deslippe, and Daniel Neuhauser}
\email{dxn@ucla.edu}
\newcommand{\lyxmathsym}[1]{\ifmmode\begingroup\def\b@ld{bold}
  \text{\ifx\math@version\b@ld\bfseries\fi#1}\endgroup\else#1\fi}

\providecommand{\tabularnewline}{\\}

\makeatother

\usepackage{babel}
\begin{document}
\maketitle
\begin{abstract}
$\mathtt{StochasticGW}$ is a code for computing accurate Quasi-Particle
(QP) energies of molecules and material systems in the GW approximation.
$\mathtt{StochasticGW}$ utilizes the stochastic Resolution of the
Identity (sROI) technique to enable a massively-parallel implementation
with computational costs that scale semi-linearly with system size,
allowing the method to access systems with tens of thousands of electrons.
We introduce a new implementation, $\mathtt{StochasticGW-GPU}$, for
which the main bottleneck steps have been ported to GPUs and which
gives substantial performance improvements over previous versions
of the code. We showcase the new code by computing band gaps of hydrogenated
silicon clusters ($\textrm{S}\textrm{i}_{\textrm{x}}\textrm{H}_{\textrm{y}}$)
containing up to 10001 atoms and 35144 electrons, and we obtain individual
QP energies with a statistical precision of better than $\pm0.03$
eV with times-to-solution on the order of minutes.
\end{abstract}

\section{Introduction}

In  recent years, predicting electronic properties of materials from
first-principles has become a key step in the materials design process,
greatly reducing laboratory time and costs by directing synthetic
efforts towards the most promising material candidates for a given
application. Properties of interest, including band gaps, ionization
potentials, and optical spectra, can be computed via electronic structure
methods implemented in commercially-available and open source software.
For excited states, post-Hartree-Fock methods, including multi-reference
configuration interaction \cite{WernerMRCI,BuenkerMRCI} and equation-of-motion
coupled-cluster methods \cite{KrylovEOMCC,StantonEOMCC}, while being
gold standards for accuracy, are only applicable to small molecules
since the computational cost of these methods grows steeply with the
number of electrons. Due to their more favorable scaling, Density
Functional Theory (DFT)-based methods \cite{EarlyDFT} have become
the industry standards for predicting ground state energies of large
molecules and materials \cite{GordonBell_2023}; however, their accuracy
is poor when used to predict Quasi-Particle (QP) energies corresponding
to excited states \cite{Martin_2004Book,Dreizler_1990Book,Aryasetiawan_TheGW}.
Excited-state methods that can be applied on top of a DFT starting
point, such as Time-Dependent (TD-)DFT \cite{EarlyTDDFT}, the GW
approximation \cite{OldGW,Aryasetiawan_TheGW,InteractElecBook}, including
its extensions using perturbation theory \cite{GWwithPT} and the
Bethe-Salpeter Equation (BSE) approach \cite{GWwithBSE}, provide
superior accuracy compared to DFT, but they are expensive to apply,
limiting excited state calculations to systems containing $\sim$10,000
electrons \cite{DelBen_LargeGWHPC,DelBen_AccelGWLeadership,MBPTBigTwist,BigGW_WEST,sc24_GW13Katoms,SC25_BGW}. 

The GW method has emerged as a robust and routinely-used tool for
computing QP energies of material systems \cite{OldGW,OldGW_1stprinciples,OldGW_eleccorr},
and GW implementations are now found in many quantum chemistry/materials
software packages \cite{VASP_liquidmetal,VASP_AIMD,QE_mainref,ABINIT_mainref,Yambo_mainref,FHI-aims_mainref,BGW_mainref,ELK_mainref,Exciting_mainref,WEST_mainref,Fiesta_mainref,MolGW_mainref,CP2K_main,SternheimerGW_mainref}.
Here, one approximates the self-energy operator, $\Sigma$, which
embodies the many-body electron exchange and correlation effects,
as the product of the single-particle Green's function, $G$, and
the screened Coulomb interaction, $W$, i.e. $\Sigma=iGW$. In common
practice, one initiates a GW calculation by first solving the Kohn-Sham
equation using a DFT method of choice to generate the starting orbitals
and energies,

\begin{eqnarray}
\left[-\frac{1}{2}\nabla^{2}+V_{ion}+V_{H}+V_{XC}^{KS}\right]\phi_{k}^{KS} & = & \varepsilon_{k}^{KS}\phi_{k}^{KS}\label{eq:KS}
\end{eqnarray}
where $V_{ion}$, $V_{H}$, and $V_{XC}$ are the ionic, Hartree,
and exchange correlation potential terms, respectively, and $\phi_{k}^{KS}$
and $\varepsilon_{k}^{KS}$ are the $k$-th orbital and energy eigenpair.
To obtain QP wavefunctions and energies, one starts by setting $\phi_{k}^{QP}=\phi_{k}^{KS}$
and $\varepsilon_{k}^{QP}=\varepsilon_{k}^{KS}$ and then solves the
analogous Dyson equation,

\begin{eqnarray}
\left[-\frac{1}{2}\nabla^{2}+V_{ion}+V_{H}+\Sigma\left(\varepsilon_{k}^{QP}\right)\right]\phi_{k}^{QP} & = & \varepsilon_{k}^{QP}\phi_{k}^{QP}\label{eq:Dyson}
\end{eqnarray}
for $\phi_{k}^{QP}$ and $\varepsilon_{k}^{QP}$ to self-consistency
\cite{GWselfconsist1,GWselfconsist2,SimpleSCGW}. For many practical
applications, it is sufficient to solve the equation in a single pass,
possibly from a pre-optimized starting point \cite{GWhybridStartPoint}.
This is referred to as the $G_{0}W_{0}$ approximation, and this is
what we use throughout the manuscript with the zero subscript omitted
for clarity.

Evaluating the self-energy operator is costly and can be tackled by
one of two strategies, broadly defined as ``deterministic'' and
``stochastic''. In deterministic GW, the overall cost is dominated
by computing the inverse dielectric $\epsilon^{-1}$ and $\Sigma$
operator matrix elements, requiring one to evaluate many integrals
and summations over valence-conduction orbital pairs; this formally
scales as $\mathcal{O}\left(N_{e}^{4}\right)$ for an $N_{e}$-electron
molecule or periodic system. Considerable efforts have been directed
towards improving this scaling: one can achieve $\mathcal{O}\left(N_{e}^{3}\log N_{e}\right)$
complexity by employing, for example, the space-time formulation and
using the Fast Fourier Transform (FFT) to transform to-and-from the
real space \cite{SpaceTimeGW_1999,SpaceTimeGW_2016,SpaceTimeGW_2018,SpaceTimeGW_2020}.
Interpolative density fitting \cite{GW_InterpDensFit} also achieves
cubic complexity, potentially with smaller prefactors than the real
space-time methods. The stochastic pseudobands approach \cite{Altman_StochPseudoBands}
can be used to reduce the size of the valence space needed to converge
matrix elements of $\Sigma$ even further, decreasing the overall
scaling to $\mathcal{O}\left(N_{e}^{2.4}\right)$.

The developments described above have spurred increasing interest
in performing large-scale GW calculations \cite{DelBen_AccelGWLeadership,sc24_GW13Katoms,SC25_BGW},
and several massively-parallel deterministic GW implementations have
been benchmarked. Zhang \emph{et al} recently demonstrated a portable
GPU implementation in the BerkeleyGW code capable of scaling efficiently
to entire exascale architectures, achieving excellent time to solution
(on the order of minutes) for the computation of quasiparticle (QP)
energies in semiconductor systems containing up to 17574 atoms in
the simulation cell \cite{SC25_BGW}. Yu and Govoni computed states
of an interface model of Si and $\textrm{S}\textrm{i}_{\textrm{3}}\textrm{N}_{\textrm{4}}$
with up to 2376 atoms and 10368 electrons on 10368 V100 GPUs in $\sim$578
minutes (summed total of $\mathtt{wstat}$ and $\mathtt{wfreq}$ steps)
using the GPU-enabled $\mathtt{WEST}$ code \cite{BigGW_WEST}. Wu
\emph{et al} reported calculations on 13824-atom, 13824-electron LiH
supercells on 449280 SW26010Pro cores in 285 s using a massively-parallel
version of $\mathtt{PWDFT}$ \cite{sc24_GW13Katoms}. Very recently,
Vetsch \emph{et al} performed non-equilibrium Green's function calculations
on hydrogen-passivated silicon nanoribbons with up to 42240 atoms
on 37600 MI250X GPUs in 42 s per iteration using a novel self-consistent
GW algorithm with domain decomposition, implemented in their $\mathsf{QuaTrEx}$
code \cite{SC25_QuaTrEx}.

For systems containing thousands of atoms or more, one can evaluate
the self-energy operator using a stochastic GW formulation at greatly
reduced cost, as detailed by our previous works \cite{BreakingLimitsGW,stochGWmolec,swiftGW,Bradbury2022,Bradbury2023_WvW,gapfilter,GWhybridStartPoint}.
Here, we briefly summarize the main features of the method. First,
we evaluate the self-energy operator in the time domain to exploit
the direct product computation of $\varSigma\left(t\right)$ from
Green's function $G$ and screened Coulomb potential $W$; we Fourier
transform the resulting $\Sigma\left(t\right)$ to $\Sigma\left(\omega\right)$
only in the final stage of the calculation. Second, we invoke the
stochastic Resolution of Identity (sRoI) \cite{Hutchinson_sROI,BreakingLimitsGW,swiftGW} and
define random orbital functions to use as bases for evaluating the
Green's function $G$ and effective polarization $W$. We then compute
the expectation values of these operators using real time propagation
and accumulate statistical averages over products of random samples.
This is the main ingredient of stochastic GW and it has the advantage
of allowing one to decouple the spatial- and time- dependence in the
six-dimensional integrals needed to evaluate $\Sigma\left(t\right)$
\cite{BreakingLimitsGW}. As a result, instead of requiring the full
space of occupied and unoccupied orbitals and energies $\left\{ \phi_{k},\varepsilon_{k}\right\} $
(which typically number in the tens of thousands or for a thousand-atom
molecule), we, in effect, evaluate $G$ and $W$ using compact sets
of stochastic linear combinations of the occupied or unoccupied orbital
space. An additional benefit of sROI is that computations in the stochastic
bases can be done independently, enabling the critical path of the
calculation to be made embarrassingly parallelizable. Third, we incorporate
sparse stochastic compression \cite{swiftGW,bradbury_deterministicfragmented-stochastic_2023,sereda_sparse-stochastic_2024,no_more_gap,param-vw,chen_mixed_2025,Allen_prb2025} in our stochastic Time-Dependent
Hartree propagation algorithm \cite{BreakingLimitsGW,LR-TDDFT} for
evaluating $W$. This enables computing components of $W$ over a
collection of randomly-chosen short segments without needing a full
spatial grid, reducing storage costs. Finally, instead of projecting
each stochastic sample onto the full set of occupied orbitals $\left\{ \phi_{k}^{KS}\right\} _{occ}$
from the preliminary DFT calculation (incurring substantial I/O and
computational costs), we filter these samples to generate occupied
stochastic orbitals. We construct a filter from a Chebyshev polynomial
expansion of the Kohn-Sham Hamiltonian \cite{Baer_ChebElecStr}. While
the filtering approach has the disadvantage of requiring many terms
to produce a sharp cutoff at the Fermi energy, we recently found that
the expansion length of the filter can be greatly decreased \cite{gapfilter}
by relaxing the expansion to have zero weights inside the band gap
(where no states are present); this, in turn, reduces the number of
matrix-vector products needed to prepare the occupied stochastic orbitals.

The above framework enables a near-linear $\mathcal{O}\left(N_{e}\log N_{e}\right)$
scaling stochastic GW algorithm, with costs dominated by performing
FFTs on the spatial grid. While development of stochastic algorithms
for computing electronic properties has lagged behind that of deterministic
ones \cite{SC25_BGW}, for QP \emph{energies} stochastic GW is well-suited
for handling large systems at a much reduced computational cost compared
to deterministic GW. Some large-scale stochastic calculations have
been performed: Vlcek \emph{et al} \cite{swiftGW} computed HOMO-LUMO
gaps of $\varGamma$-point Diamond and Silicon supercells containing
up to 2744 atoms and 10976 electrons in under 2000 core hours on an
HPC cluster containing 144 nodes with 1728 Intel Xeon E5-2680v3@2.5
GHz processors. More recently, Brooks \emph{et al} \cite{MBPTBigTwist}
used $\mathtt{StochasticGW}$ to compute twist-induced localized Moire
states of bi-layer phosphorene sheets containing up to 2708 atoms
and 13540 electrons. Both of these calculations were performed with
a CPU-only version of the code without gapped-filtering. In this paper,
we assimilate the ideas described above in a new GPU-accelerated version
of $\mathtt{StochasticGW}$ which we showcase by computing QP energies
of clusters containing upwards of 10001 atoms and 35144 electrons
on $\sim1000$ GPUs with times-to-solution of minutes. 

\section{Implementation Details}

\subsection{Algorithmic Overview}

\begin{figure}
\includegraphics[bb=0bp 100bp 900bp 385bp,scale=0.6]{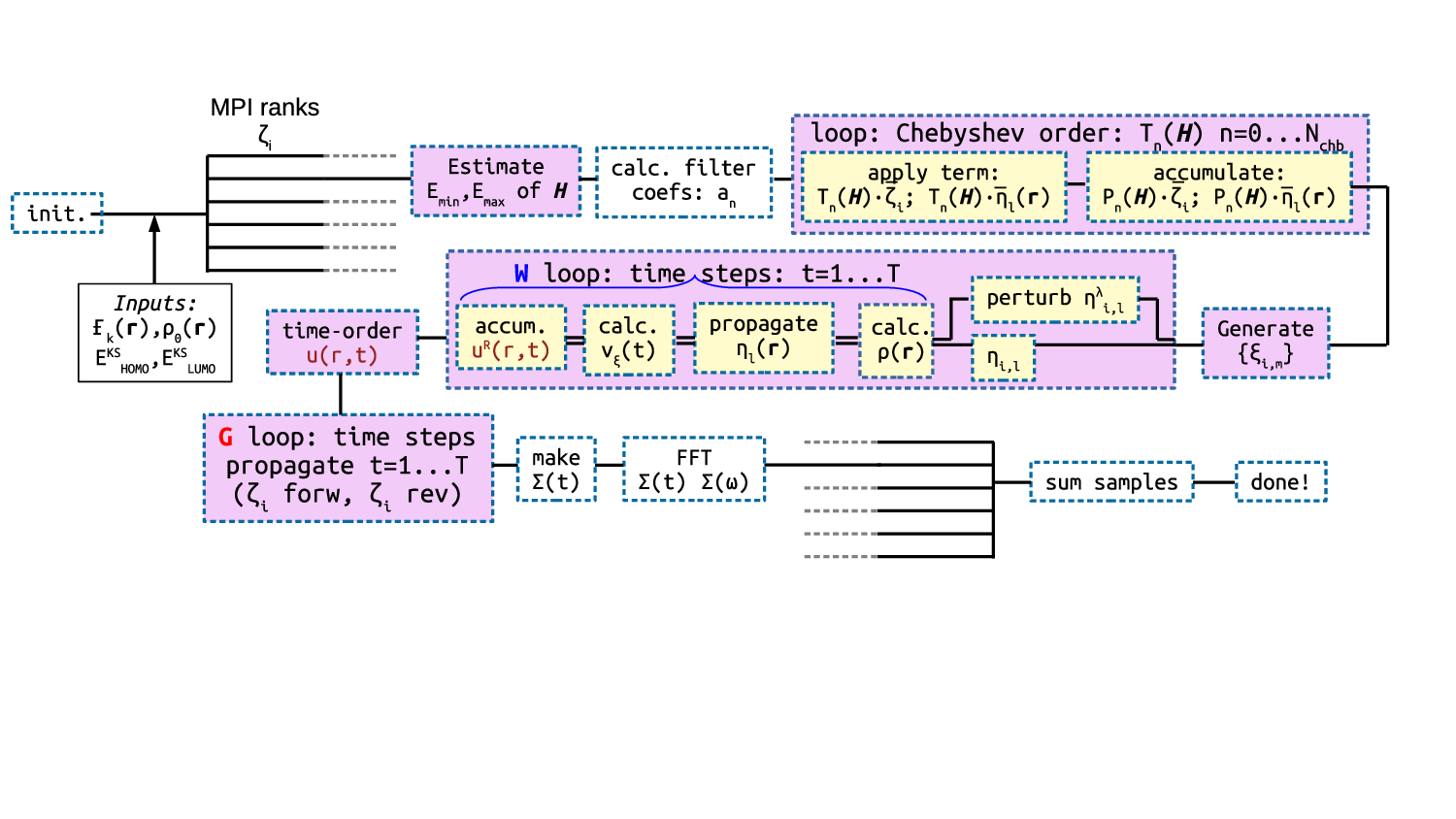}

\caption{\label{fig:sGWalgo}Block diagram of the main steps of the StochasticGW
algorithm. Each MPI rank performs the same operations on a different
set of data (see text for details). Steps enclosed in shaded boxes
have been ported to GPUs.}
\end{figure}

Our stochastic GW implementation is similar to that described previously
\cite{swiftGW} and with the inclusion of the gapped filtering \cite{gapfilter}
technique. Here, we only summarize the key components of the algorithm;
see the earlier works for a more detailed explanation of the methodology.
A block diagram, shown in Figure \ref{fig:sGWalgo}, depicts the major
portions of the code.

The $\mathtt{StochasticGW}$ code requires as inputs: 1) coordinates
of the atoms, 2) a pseudopotential for each atomic type, 3) the ground
state density $\rho_{0}\left(\mathbf{r}\right)$, 4) estimates of
the energies of the highest occupied and lowest unoccupied molecular
orbitals ($E_{HOMO}^{KS}$ and $E_{LUMO}^{KS}$, respectively), and
5) a spatial orbital $\phi_{k}$ for which the quasiparticle energy
$\varepsilon_{k}^{QP}$ is desired. Items 1) - 3) are needed to construct
the Kohn-Sham Hamiltonian internally in $\mathtt{StochasticGW}$;
item 4) defines the cutoff region of the gapped filter. We obtain
items 3) - 5) from a preliminary DFT calculation.

We begin by constructing the spectral filter to apply the Heaviside
operator. The Heaviside operator ($\Theta$) expanded in Chebyshev
polynomials $T_{n}$ in the Hamiltonian $\hat{H}$, which can be written
as:

\begin{equation}
\Theta\left(\mu-\hat{H}\right)\simeq\sum_{n=0}^{N_{chb}}a_{n}T_{n}\left(\hat{H}\right)\label{eq:HeavisideExpn}
\end{equation}
where $\mu$ is the chemical potential, $a_{n}$ are expansion coefficients,
and $N_{chb}$ is the maximum degree of Chebyshev polynomial needed
to build the filter. One requires the upper and lower spectral bounds
of the Kohn-Sham Hamiltonian $\hat{H}^{KS}$ to shift-and-scale the
eigenvalue spectrum into the interval $\left[-1,1\right]$; there
are various schemes to obtain these bounds but we find that one of
the simplest, a shifted power iteration, works well for this purpose.
In the gapped-filtering method, rather than applying the cutoff of
filter $\Theta$ at a specific value of $\mu$ we instead apply it
over the band gap containing $\mu$, i.e. over $E_{HOMO}^{KS}\leq\mu\leq E_{LUMO}^{KS}$,
so we must also map the energies $E_{HOMO}^{KS}$ and $E_{LUMO}^{KS}$
into $\left[-1,1\right]$. With these we compute the $N_{chb}$ filter
coefficients $a_{n}$ as described previously \cite{gapfilter}.

Next, we generate $N_{\zeta}$ random ``white-noise'' start orbitals,
$\left|\bar{\zeta}_{i}\left(\mathbf{r},t=0\right)\right\rangle $,
for the stochastic realization of $G$; we dub these the Monte Carlo
(MC) samples. For each, we also generate $N_{\eta}$ additional white-noise
orbitals, $\left|\bar{\eta}_{i,\ell}\left(\mathbf{r},t=0\right)\right\rangle $,
needed to calculate the action of the time-dependent effective interaction
operator $W\left(t\right)$ on a vector related to each $\left|\bar{\zeta}_{i}\right\rangle $.
We apply the Heaviside filter to both sets of $\left\{ \bar{\zeta}_{i}\right\} $
and $\left\{ \bar{\eta}_{i,\ell}\right\} $ orbitals in order to project
them onto random linear combinations of the occupied orbital subspace
$\left\{ \phi_{k}^{KS}\right\} _{occ}$. We denote these ``filtered''
orbitals as $\left\{ \zeta_{i}\right\} $ and $\left\{ \eta_{i,\ell}\right\} $
(without the overhead bar).

Subsequently, we evaluate the diagonal time-dependent self-energy
matrix element, $\left\langle \phi_{k}\left|\Sigma\left(t\right)\right|\phi_{k}\right\rangle $,
in two phases. In the first, we use linear-response time-dependent
Hartree \cite{BreakingLimitsGW} to compute the action of the retarded
polarization interaction $W^{R}$ on the occupied states. Algorithmically,
we create a perturbed copy of each $\eta_{i,\ell}$ (denoted $\eta_{i,\ell}^{\lambda}$),
and propagate both perturbed and unperturbed copies in time under
the action of the time-dependent Hamiltonian. We accumulate the causal
response function, $u^{R}\left(\mathbf{r},t\right)$, as the difference
of the perturbed and unperturbed time-dependent potentials, and we
time-order \cite{TimeOrder} the accumulated $u^{R}\left(\mathbf{r},t\right)$
to produce the effective polarization potential $u\left(\mathbf{r},t\right)$.
Note that instead of accumulating $u^{R}\left(\mathbf{r},t\right)$
in the basis of $\left\{ \eta_{i,\ell}\right\} $ (which requires
keeping a copy of $u^{R}\left(\mathbf{r},t\right)$ on the full spatial
grid for each $\eta_{i,\ell}$), we reduce the storage requirement
by projecting $u^{R}\left(\mathbf{r},t\right)$ onto a set of $N_{\xi}$
randomly chosen functions $\left|\xi_{i,m}\right\rangle $ on short,
fragmented segments of the spatial grid and accumulate $u^{R}\left(\mathbf{r},t\right)$
in the $\left\{ \xi_{i,m}\right\} $ basis.

In the second phase, we evaluate the action of the Green's function
$iG$ on each $\bar{\zeta}_{i}$. Numerically, we compute $\zeta_{i}\left(t\right)$
by propagating the filtered $\zeta_{i}$ (representing a random linear
combination of occupied states) backwards in time while simultaneously
propagating its orthogonal complement (representing a random linear
combination of unoccupied states) forwards in time. This represents
the stochastic realization of Green's function as a time correlation
function,

\begin{eqnarray}
iG\left(\mathbf{r},\mathbf{r^{\prime}},t\right) & = & \frac{1}{N_{\zeta}}\sum_{\zeta}\zeta\left(\mathbf{r},t\right)\bar{\zeta}\left(\mathbf{r^{\prime}}\right).\label{eq:GreensStoch}
\end{eqnarray}
Having obtained $u\left(\mathbf{r},t\right)$ and $\zeta\left(\mathbf{r},t\right)$
from the first and second phases, respectively, the diagonal self-energy
matrix element for orbital $\phi_{k}$ becomes

\begin{eqnarray}
\left\langle \phi_{k}\left|\Sigma\left(t\right)\right|\phi_{k}\right\rangle  & = & \int\int\phi_{k}\left(\mathbf{r}\right)iG\left(\mathbf{r},\mathbf{r}^{\prime},t\right)W\left(\mathbf{r},\mathbf{r^{\prime}},t\right)\phi_{k}\left(\mathbf{r^{\prime}}\right)d\mathbf{r}d\mathbf{r^{\prime}}\label{eq:SigmaElement}\\
 & = & \frac{1}{N_{\zeta}}\sum_{\zeta}\int\int\phi_{k}\left(\mathbf{r}\right)\zeta\left(\mathbf{r},t\right)W\left(\mathbf{r},\mathbf{r^{\prime}},t\right)\bar{\zeta}\left(\mathbf{r^{\prime}}\right)d\mathbf{r}d\mathbf{r^{\prime}}\nonumber \\
 & = & \frac{1}{N_{\zeta}}\sum_{\zeta}\int\phi_{k}\left(\mathbf{r}\right)\zeta\left(\mathbf{r},t\right)u\left(\mathbf{r},t\right)d\mathbf{r}.\nonumber 
\end{eqnarray}
We then compute the the frequency-resolved self-energy matrix element
$\left\langle \phi_{k}\left|\Sigma\left(\omega\right)\right|\phi_{k}\right\rangle $
from the time-dependent form via discrete Fourier transform, and we
obtain quasi-particle energy, $\varepsilon_{k}^{QP}$, by solving

\begin{eqnarray}
\varepsilon_{k}^{QP} & = & \varepsilon_{k}^{KS}+\left\langle \phi_{k}\left|X+\Sigma\left(\omega=\varepsilon_{k}^{QP}\right)-V_{XC}\right|\phi_{k}\right\rangle \label{eq:Eqp}
\end{eqnarray}
where $X$ is the sROI \cite{BreakingLimitsGW,swiftGW} realization of the Fock exchange operator in
the basis of $\left\{ \eta_{i,\ell}\right\} $ and all other quantities
have been previously defined.

\subsection{Scaling of the method}

A key aim of our stochastic GW formulation is to achieve computational
scaling that grows slowly, ideally linearly, with respect to system
size. The most numerically intensive portions of the algorithm apply
matrix-vector products repeatedly to the set of $\left\{ \eta_{i,\ell}\right\} $
during the filtering and propagation phases. Here, one applies either
the Kohn-Sham Hamiltonian, $\hat{H}^{KS}$, or the evolution operator,
$e^{-i\hat{H}(t)\Delta t}$, to a set of vectors with each having
length $N_{g}=N_{x}N_{y}N_{z}$. We apply matrix-vector products in
a Fourier grid representation whereby FFT pairs switch between position
and momentum representations where the potential energy and kinetic
energy operators are diagonal, respectively. Applying the individual
kinetic and potential energy operators scales as $\mathcal{O}\left(N_{g}\right)$,
but the cost of each Hamiltonian/evolution operation is dominated
by the $\mathcal{O}\left(N_{g}\log_{2}N_{g}\right)$ FFT cost. Thus,
to filter the $N_{\zeta}N_{\eta}$ starting orbitals, one applies
a length $N_{chb}$ filter at a cost of $\mathcal{O}\left(N_{\zeta}N_{\eta}N_{chb}N_{g}\log_{2}N_{g}\right)$
operations. Likewise, propagating the full set of $\left\{ \eta_{i,\ell}\right\} $
for $N_{\tau}$ time steps has a cost scaling as $\mathcal{O}\left(N_{\zeta}N_{\eta}N_{\tau}N_{g}\log_{2}N_{g}\right)$.
Accumulating $u^{R}\left(\mathbf{r},t\right)$ costs $\mathcal{O}\left(N_{\zeta}N_{\xi}N_{\tau}N_{g}f_{g}\right)$,
where $f_{g}$ is the fractional length of each of the fragmented
stochastic functions $\left\{ \xi_{i,m}\right\} $ relative to the
full spatial grid length, $N_{g}$.

For tackling quasi-particle energies of large molecules it is important
to consider the dependence of each parameter on system size. The number
of MC samples, $N_{\zeta}$, and the number of occupied stochastic
orbitals, $N_{\eta}$ determine the statistical accuracy of the QP
energies and do not increase with system size ($N_{\eta}$ actually
decreases with increasing system size due to self-averaging). The
number of time steps, $N_{\tau}$, determines the energy resolution
of $\Sigma\left(\omega\right)$ and is also independent of system
size. The number of grid points, $N_{g}$, while cubic in dimension
($N_{g}=N_{x}N_{y}N_{z}$), grows linearly overall with system size
due to spatial packing of atoms in 3-dimensional space. For accumulating
$u^{R}\left(\mathbf{r},t\right)$, the statistical error does increase
with the ratio $\frac{N_{g}}{N_{\xi}}.$ This means that one must
simultaneously increase the number of stochastic segments $N_{\xi}$
as the grid size $N_{g}$ increases to prevent growth of errors. However,
in the sparse stochastic basis, the cost increase from requiring larger
$N_{\xi}$ can be offset by decreasing the fractional length $f_{g}$
of each segment $\left\{ \xi_{i,m}\right\} $ (i.e. by using more
$\xi$ vectors but making them ``sparser'') \cite{swiftGW}. Finally,
the number of Chebyshev coefficients, $N_{chb}$, needed to fit the
gapped filter, depends on the width of the band gap relative to the
spectral width of the Kohn-Sham Hamiltonian. The value of $N_{chb}$
needed to accurately fit the filter does increase with system size
due to larger spectral width of $\hat{H}^{KS}$, but this can be mitigated
by applying a kinetic energy cutoff. In summary, as long as care is
taken to manage the growth of the $N_{\xi}$ and $N_{chb}$ parameters
accordingly, one can achieve near-linear scaling with system size
in stochastic GW calculations.

\subsection{GPU implementation}

The original $\mathtt{StochasticGW}$ code (through v.2.0) is written
in Fortran 90 and parallelized using Message Passing Interface (MPI).
A key feature of $\mathtt{StochasticGW}$ is that the $N_{\zeta}$
Monte Carlo samples can be processed independently of one another,
resulting in embarrassing parallelism over large portions of the algorithm.
Additionally, the code contains an option to extend the MPI-level
parallelism over the $N_{\eta}$ occupied stochastic functions at
the cost of an additional call to $\mathsf{mpi}\textrm{\_}\mathsf{allreduce()}$
at each time step (needed to compute the time-dependent density $\rho\left(\mathbf{r},t\right)$).
In the original implementation, operations over grid points are performed
in serial. For systems containing thousands of atoms or more, the
grids are large enough that these operations become significant serial
bottlenecks, motivating us to develop a GPU port to handle them in
parallel.

In the GPU implementation, we retain the idea of processing each of
the $N_{\zeta}$ MC samples with a separate MPI rank, but the $N_{\eta}$
occupied stochastic functions per sample reside on the same MPI rank
so that MPI calls are not needed at each time step to evaluate the
time-dependent density $\rho\left(\mathbf{r},t\right)$. The GPU code
utilizes kernels written using OpenACC directives and calls to specialized
libraries ($\mathsf{cuRAND}$ and $\mathsf{cuFFT}$) when needed.
To maximize efficency, attention must be given to minimize the amount
of data transferred between the host CPU and each GPU and to organize
the computational workload to expose as much parallelism to the GPU
as possible. To this end, we performed several optimizations.

First, we structured the arrays containing the stochastic orbitals
with multi-indices so that each kernel can process the orbitals in
single-instruction-multiple-data (SIMD) fashion. Many of the operations
in the filtering and propagation cycles, such as applying the kinetic
and potential energy operators, are simple element-wise array multiplications
which are highly vectorizable on GPU hardware. In each case which
follows, we construct the multi-index arrays and then offload them
once onto a single GPU, retrieving the result only after the full
set of filtering or propagation iterations. For the filtering cycle,
this means that on each MPI rank we pack the $\bar{\zeta}_{i}$ associated
with rank $i$ along with its set of $\left\{ \bar{\eta}_{i,\ell}\right\} ;\ell=1\ldots N_{\eta}$
orbitals into an array of size $\left(N_{g}\times N_{\eta}+1\right)$.
For the propagation cycle involving the set of $\left\{ \eta_{i,\ell}\right\} $,
the perturbed and unperturbed copies can be processed in parallel,
so we pack both copies into an array of size $\left(N_{g}\times N_{\eta}\times2\right)$.
The MC sample $\zeta_{i}$ cannot be propagated in parallel with the
$\left\{ \eta_{i,\ell}\right\} $ here since the former depends on
the effective polarization potential $u\left(\mathbf{r},t\right)$
resulting from the $\left\{ \eta_{i,\ell}\right\} $ propagation cycle.
However, since reverse-time propagation of $\zeta_{i}$ and forward-time
propagation of its orthogonal complement are operationally identical
(other than a difference in sign), we can pack these functions into
an array of size $\left(N_{g}\times2\right)$ and achieve a parallel
performance boost for propagation of $\zeta_{i}$ as well.

Not all operations in the filtering and propagation steps are trivial
to vectorize. Normalizations appear periodically in each of the filtering,
propagation, and spectral estimation stages; each requires summing
over values defined over $N_{g}$ grid points. For instance, for normalizations
performed in the $\left\{ \eta_{i,\ell}\right\} $ propagation cycle,
at most only $2N_{\eta}$ operations can be performed in parallel
instead of $2N_{\eta}N_{g}$. For the largest systems in this work,
$N_{\eta}\sim8$ while $N_{g}\sim1.6\times10^{8}$, meaning that the
benefits of having many parallel threads are largely lost in each
normalization kernel call. To solve this, we divided the $N_{g}$
grid points into short segments of length $L$. This allows us to
parallelize sums over grid points over $\frac{N_{g}}{L}$ threads
at the cost of having to perform an atomic add by each thread after
the sum over each segment has been accumulated. The optimal value
of $L$ is hardware dependent; on NVIDIA A100 GPUs we achieved the
best performance with $L\sim256$. In this manner, we achieve an overall
parallelism of up to $2N_{\eta}\frac{N_{g}}{L}$ threads in normalization
calls.

Second, the main computations needed to accumulate $u\left(\mathbf{r},t\right)$
have also been ported to the GPU. We generate the $\left\{ \xi_{i,m}\right\} $
basis via calls to the $\mathsf{cuRAND}$ library, and we compute
the overlaps $\left\langle \xi\mid u^{R}\left(t\right)\right\rangle $
on-the-fly during the $\left\{ \eta_{i,\ell}\right\} $ propagation
phase in a segmented fashion similar to the one described above for
normalization. Here, we multiply two arrays of sizes $\left(N_{\xi}\times N_{g}f_{g}\right)$
and $\left(N_{g}f_{g}\times2\right)$ parallelized over $2N_{\xi}\frac{N_{g}f_{g}}{L}$
threads, where each $\xi$ function is processed in segments of length
$L=32$, and the factor of 2 again arises from performing the unperturbed-$\eta$
and perturbed-$\eta$ propagations in parallel. Finally, we perform
the time-ordering operation to convert the resulting $u^{R}\left(\mathbf{r},t\right)$
to $u\left(\mathbf{r},t\right)$ by calling the $\mathsf{cuFFT}$
library before and after an OpenACC kernel used for performing the
complex conjugation step.

\subsection{Utilities}

The newest (3.0) version of $\mathtt{StochasticGW}$ is freely available
\cite{sGWcode} on GitHub and includes several utilities to aid researchers
in preparing inputs for the code: 

The $\mathtt{dft2sgw}$ utility reads and preprocesses results from
a preliminary DFT calculation. This utility requires a DFT output
file and a set of $\mathsf{.cube}$ files as input; $\mathtt{dft2sgw}$
prepares an input file, $\mathsf{sgwinp.txt}$, containing atomic
coordinates, HOMO and LUMO energies (for gapped filtering), the spatial
charge density, and a requested set of orbitals for the system of
interest. $\mathtt{dft2sgw}$ also has a functionality, enabled via
the $\mathtt{FFTW}$ \cite{FFTW_main} library, to up- or down-sample
the orbital/density spatial grid from the preliminary DFT calculation
in case a different grid for the stochastic GW step is desired. The
utility currently supports Quantum ESPRESSO \cite{QE_mainref,QE_exascale},
the Real-Space Multigrid ($\mathtt{RMG}$)- \cite{RMGDFT-1996,RMGDFT-2007}
DFT code, and $\mathtt{CP2K}$ \cite{CP2K_main} (but note that $\mathtt{StochasticGW}$
does not yet include pseudopotential support for $\mathtt{CP2K}$).

$\mathtt{StochasticGW}$ also includes two utilities, $\mathtt{plotfilter.py}$
and $\mathtt{plotorbital.py}$, which use the $\mathtt{MatPlotLib}$
\cite{MatPlotLib_main} python package to generate plots related to
stochastic GW calculations:

The $\mathtt{plotfilter.py}$ utility plots the filter and depicts
the log of the magnitudes of the filter coefficients and is useful
for checking the quality of the Chebyshev expansion of the filter. 

The $\mathtt{plotorbital.py}$ utility visualizes the atomic coordinates,
spatial orbitals, and charge density contained in the file sgwinp.txt;
this feature allows one to quickly select orbitals of interest for
a subsequent $\mathtt{StochasticGW}$ calculation.

\section{Numerical Experiments}

\begin{figure}
\includegraphics[bb=0bp 100bp 612bp 600bp,scale=0.5]{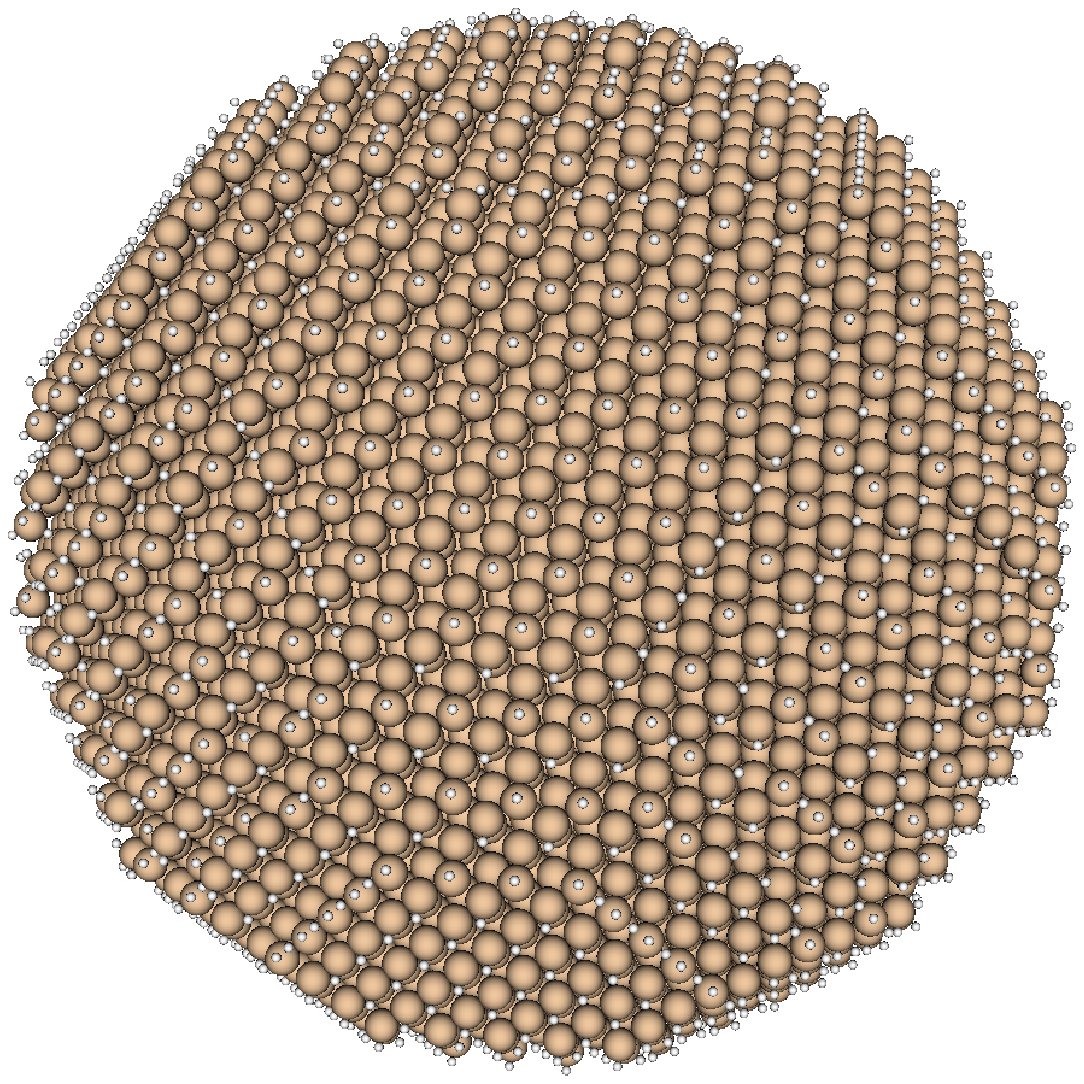}

\caption{\label{fig:Si8381H1620}$\textrm{S}\textrm{i}_{\textrm{8381}}\textrm{H}_{\textrm{1620}}$
cluster. Silicon and hydrogen atoms shown as brown and white spheres,
respectively.}
\end{figure}

We now test our implementation of $\mathtt{StochasticGW}$ by computing
QP energies of a series of non-periodic nanoclusters, $\textrm{S}\textrm{i}_{\textrm{293}}\textrm{H}_{\textrm{172}}$,
$\textrm{S}\textrm{i}_{\textrm{705}}\textrm{H}_{\textrm{300}}$, $\textrm{S}\textrm{i}_{\textrm{5031}}\textrm{H}_{\textrm{1172}}$,
$\textrm{S}\textrm{i}_{\textrm{7745}}\textrm{H}_{\textrm{1572}}$,
$\textrm{S}\textrm{i}_{\textrm{8381}}\textrm{H}_{\textrm{1620}}$.
We constructed each cluster from a uniformly expanded silicon superlattice
of size $15\times15\times15$ using the experimental unit cell parameter
for silicon ($a=b=c=10.26$ Bohr) corresponding to the diamond cubic
structure with an eight-atom unit cell \cite{si_unit}. We then shifted
the coordinate origin to the geometric center of the superlattice
and applied a spherical truncation, retaining only Si atoms within
20-70 Bohr of the origin. The truncated cluster is passivated with
hydrogen atoms to saturate dangling bonds, and the resulting structure
is relaxed to its equilibrium geometry using the MMFF94 force field
\cite{mmff94} as implemented in Open Babel software \cite{open_babel}.
Figure \ref{fig:Si8381H1620} depicts the largest cluster, $\textrm{S}\textrm{i}_{\textrm{8381}}\textrm{H}_{\textrm{1620}}$.

We performed the initial periodic DFT calculations to generate the
orbitals and charge densities using the $\mathtt{RMG}$ \cite{RMGDFT-1996,RMGDFT-2007}
DFT code. The DFT Hamiltonian uses the GGA PBE exchange-correlation
functional with Troullier-Martins \cite{TroullierMartinsPP} norm-conserving
pseudo-potentials. For each system, we performed the calculation on
the Gamma k-point in a primitive cubic cell with isotropic sampling.
Each cluster is separated from its periodic image by a vacuum layer
of 11-17 bohr. The initial DFT step provides the energy estimates
$E_{HOMO}^{KS}$ and $E_{LUMO}^{KS}$ used to define the gapped filter
for the GW step. Cell and grid parameters, along with HOMO and LUMO
energies, are listed in Table \ref{tab:DFTresults}.

\begin{table}
\caption{\label{tab:DFTresults}Details of preliminary DFT calculations on
each cluster, including numbers of electrons ($N_{e}$), points in
the spatial grid ($N_{g}$), grid spacing ($\Delta_{g}$, bohr), along
with resulting HOMO and LUMO energies and band gaps (eV).}

\begin{tabular}{ccccccc}
\hline 
System & $N_{e}$ & $N_{g}$ & $\Delta_{g}$ & $E_{HOMO}^{KS}$ & $E_{LUMO}^{KS}$ & KS Band Gap\tabularnewline
\hline 
$\textrm{S}\textrm{i}_{\textrm{293}}\textrm{H}_{\textrm{172}}$ & 1344 & $128^{3}$ & 0.4429 & -3.259  & -1.526  & 1.733 \tabularnewline
$\textrm{S}\textrm{i}_{\textrm{705}}\textrm{H}_{\textrm{300}}$ & 3120 & $128^{3}$ & 0.5167 & -1.928  & -0.457  & 1.471 \tabularnewline
$\textrm{S}\textrm{i}_{\textrm{5031}}\textrm{H}_{\textrm{1172}}$ & 21296 & $256^{3}$ & 0.5000 & -1.705  & -0.776  & 0.928 \tabularnewline
$\textrm{S}\textrm{i}_{\textrm{7745}}\textrm{H}_{\textrm{1572}}$ & 32552 & $256^{3}$ & 0.5400 & -0.986  & -0.121  & 0.865 \tabularnewline
$\textrm{S}\textrm{i}_{\textrm{8381}}\textrm{H}_{\textrm{1620}}$ & 35144 & $256^{3}$ & 0.5600 & -1.095  & -0.245  & 0.851 \tabularnewline
\hline 
\end{tabular}
\end{table}

\begin{figure}
\includegraphics{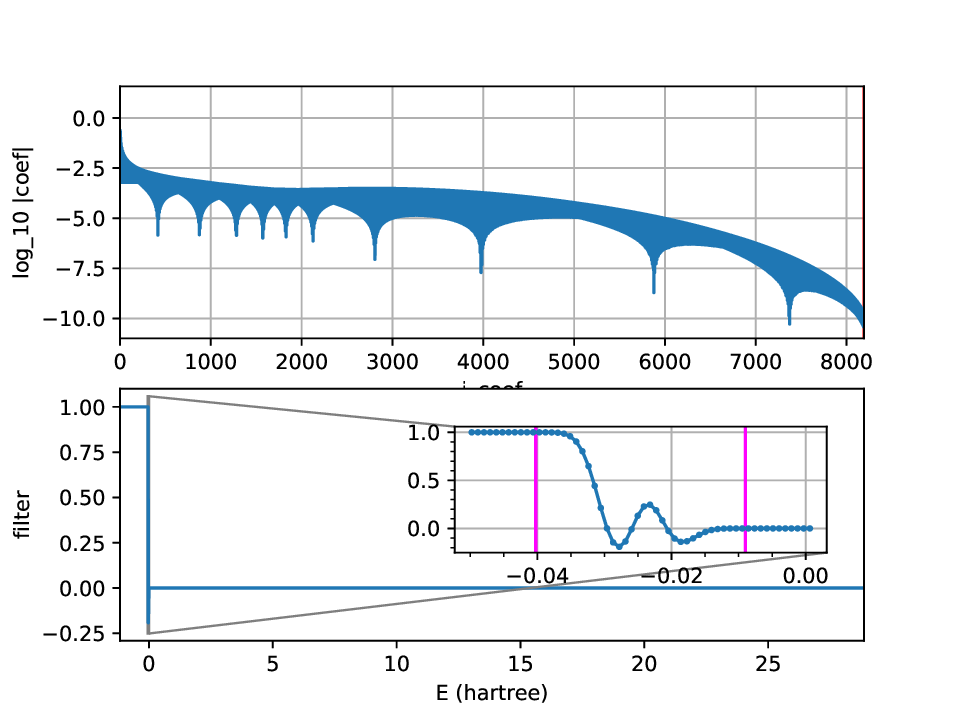}

\caption{\label{fig:filter}(Top) plot of the log of the absolute magnitudes
of Chebyshev coefficients used to construct the gapped filter for
the $\textrm{S}\textrm{i}_{\textrm{8381}}\textrm{H}_{\textrm{1620}}$
cluster. (Bottom) Reconstructed filter, where the inset shows an expansion
of the region of the band gap. The purple vertical lines in the inset
indicate the positions of $E_{HOMO}^{KS}$ and $E_{LUMO}^{KS}$.}
\end{figure}

We then used $\mathtt{StochasticGW}$ to compute $\varepsilon_{QP}$
for the HOMO and LUMO orbitals of each system. The Kohn-Sham Hamiltonian
in $\mathtt{StochasticGW}$ uses the same pseudopotentials and grids
as the previous DFT step; here, we employ the PBE functional \cite{PBEfunctional}
as implemented in the LibXC \cite{LibXC_mainref} library and apply
an energy cutoff of 28 Hartrees to the kinetic energy operator. In
the filtering step, for the largest system we studied, $\textrm{S\ensuremath{i_{8381}H_{1620}}}$,
the energy difference $E_{LUMO}^{KS}-E_{HOMO}^{KS}$ is $\sim0.11\%$
the full spectral range of $\hat{H}^{KS}$. Even though the cutoff
is spread over the full band gap, it is still sharp enough to require
8192 Chebyshev terms to reduce the Gibbs oscillations to negligible
levels outside of the band gap (Figure \ref{fig:filter}). While this
filter length is an order of magntude larger than that used in our
recent calculation on the napthalene molecule ($N_{chb}=450$) \cite{gapfilter},
it is still less than lengths required in earlier calculations performed
on much smaller systems without gapped filtering ($N_{chb}\sim18000$)
\cite{stochGWmolec}.

For each calculation we averaged 1024 Monte Carlo samples which is
sufficient to achieve a statistical error better than 0.03 eV for
all QP energies. The number of time steps, $N_{\tau}$, are controlled
internally by the energy-broadening parameter, $\gamma$, which we
apply when Fourier transforming the self-energy element from the time
domain to the frequency domain,

\begin{equation}
\left\langle \phi_{k}\left|\Sigma\left(\omega\right)\right|\phi_{k}\right\rangle =\int\left\langle \phi_{k}\left|\Sigma\left(t\right)\right|\phi_{k}\right\rangle e^{-\frac{\gamma^{2}t^{2}}{2}}e^{i\omega t}dt.\label{eq:time2freq}
\end{equation}
We use a time step size of $\Delta t=0.05E_{h}^{-1}\hbar$ in the
split-operator propagation of the orbitals. The number of time steps
to obtain a desired energy resolution is $N_{\tau}\approx\frac{3}{\gamma\cdot\Delta t}$;
for all calculations in this work, we set $\gamma=0.06E_{h}\hbar^{-1}$
which yields a propagation length of $N_{\tau}=1000$ time steps over
50 atomic time units. Numbers of occupied stochastic orbitals ($N_{\eta}$)
and segmented stochastic functions ($N_{\xi}$) along with the fractional
grid lengths ($f_{g}$) for the latter are chosen at values similar
to those in previous works \cite{stochGWmolec,swiftGW,MBPTBigTwist}.
Input parameters are summarized in Table \ref{tab:sGWparams}. All
calculations were run on 256 GPU nodes of NERSC-Perlmutter; each node
contains one AMD EPYC 7763 processor running at 2.5 GHz and 4 NVIDIA
A100 GPUs.

\begin{table}
\caption{\label{tab:sGWparams}Parameters used in stochastic GW calculations.}

\begin{tabular}{ccc}
\hline 
Description & Parameter & Value\tabularnewline
\hline 
Number of Monte Carlo samples & $N_{\zeta}$ & 1024\tabularnewline
Number of occupied stochastic orbitals & $N_{\eta}$ & 8\tabularnewline
Number of segmented stochastic functions & $N_{\xi}$ & 10000\tabularnewline
Grid fraction of each segmented function & $f_{g}$ & 0.003\tabularnewline
Number of Chebyshev polynomials in filter & $N_{chb}$ & 8192\tabularnewline
Damping parameter (Hartrees) & $\gamma$ & 0.06\tabularnewline
Kinetic energy cutoff (Hartrees) & $E_{cut}^{k}$ & 28.0\tabularnewline
\hline 
\end{tabular}
\end{table}

\begin{figure}
\includegraphics[scale=0.8]{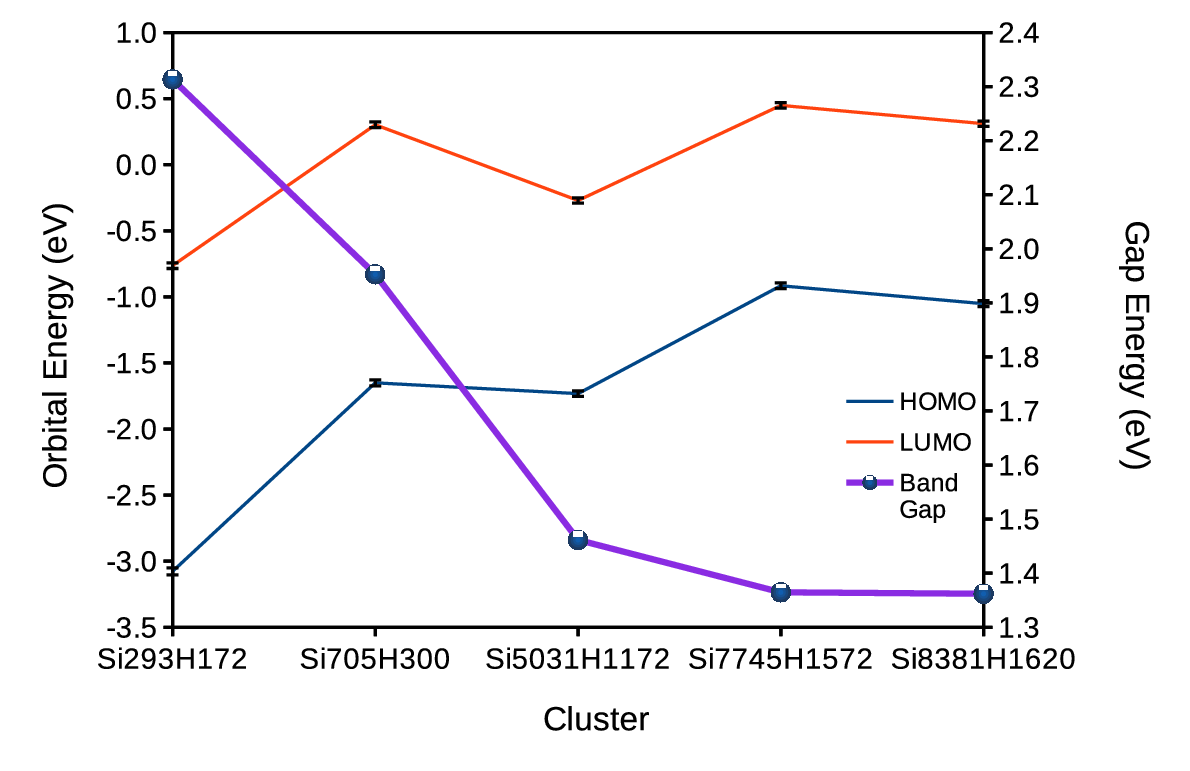}

\caption{\label{fig:BandGapPlot}QP orbital energies and band gaps for each
cluster. }
\end{figure}

Figure \ref{fig:BandGapPlot} plots the QP energies of the HOMO and
LUMO and their difference for each system; these values are also listed
in Table \ref{tab:sGWresults}. The statistical errors in the MC energies
are shown as error bars in the HOMO and LUMO traces and are small
compared to the magnitudes of the energies. Moreover, comparing the
bandgaps across the five clusters, the bandgaps show convergent behavior
towards $\sim1.36$ eV, suggesting that the largest clusters are approaching
the bulk limit for our choice of density functional and pseudopotential.

Table \ref{tab:sGWresults} also lists wall times for all calculations.
The two smaller clusters have comparable times of $800\pm40$ s and
the larger three clusters have timings of $2700\pm120$ s, where the
main factor behind the difference is the size of the spatial grid
($128^{3}$ vs $256^{3}$ for the smaller and larger clusters, respectively).
For a given spatial grid, one expects an increase in runtime for the
larger systems for two reasons: first, the potential energy terms
containing the pseudopotential contribution must be applied via atomic
operations on grid points where pseudopotentials for neighboring atoms
overlap. Second, the higher density of states in larger systems causes
the spectral range estimation to converge more slowly. However, these
differences are smaller than the variation in performancein MPI and
I/O operations that occur over the large scale of these runs.

\begin{table}
\caption{\label{tab:sGWresults}Results of stochastic GW calculations on each
cluster including quasiparticle energies and band gaps (eV) and calculation
wall times (s).}

\begin{tabular}{cccccc}
\hline 
System & $E_{HOMO}^{QP}$ & $E_{LUMO}^{QP}$ & QP Band Gap & $t_{HOMO}^{wall}$ & $t_{LUMO}^{wall}$\tabularnewline
\hline 
$\textrm{S}\textrm{i}_{\textrm{293}}\textrm{H}_{\textrm{172}}$ & $-3.077\pm0.027$ & $-0.764\pm0.022$ & 2.313 & 836 & 770\tabularnewline
$\textrm{S}\textrm{i}_{\textrm{705}}\textrm{H}_{\textrm{300}}$ & $-1.650\pm0.023$ & $0.302\pm0.022$ & 1.953 & 835 & 788\tabularnewline
$\textrm{S}\textrm{i}_{\textrm{5031}}\textrm{H}_{\textrm{1172}}$ & $-1.732\pm0.021$ & $-0.271\pm0.020$ & 1.462 & 2609 & 2617\tabularnewline
$\textrm{S}\textrm{i}_{\textrm{7745}}\textrm{H}_{\textrm{1572}}$ & $-0.916\pm0.022$ & $0.449\pm0.021$ & 1.365 & 2702 & 2688\tabularnewline
$\textrm{S}\textrm{i}_{\textrm{8381}}\textrm{H}_{\textrm{1620}}$ & $-1.052\pm0.023$ & $0.310\pm0.029$ & 1.362 & 2669 & 2812\tabularnewline
\hline 
\end{tabular}
\end{table}

\begin{figure}
\includegraphics[scale=0.8]{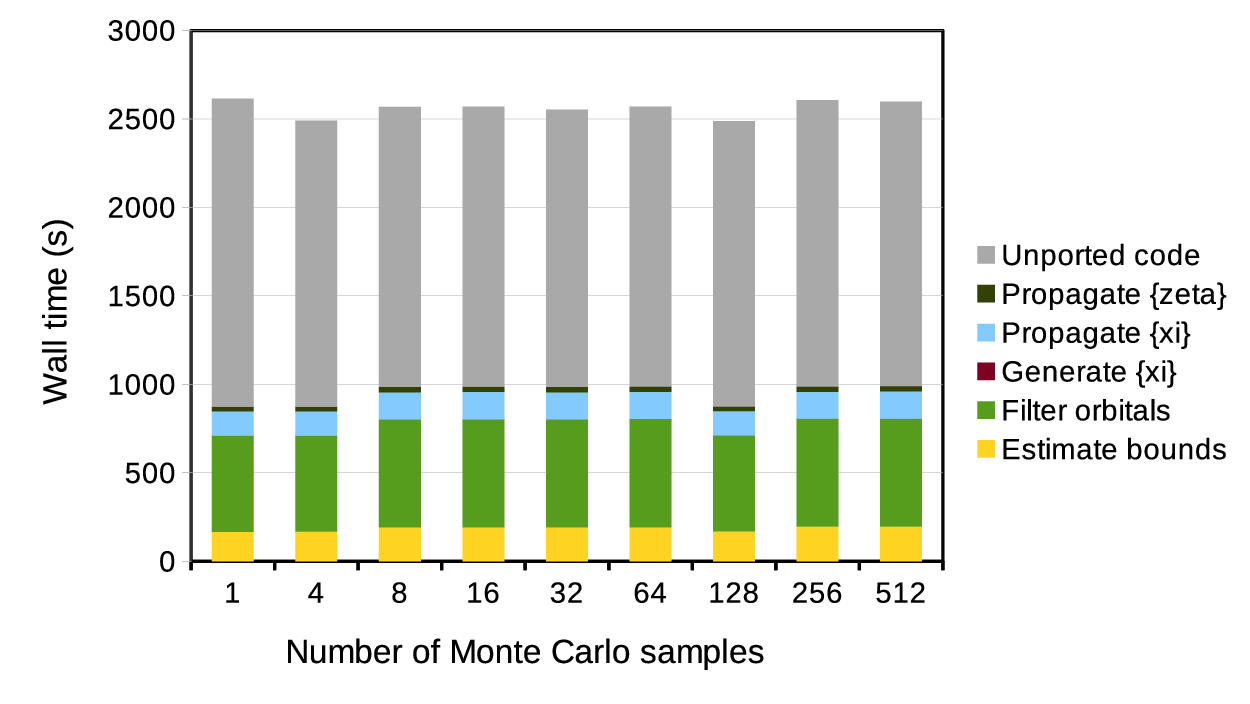}

\caption{\label{fig:BigTimings}Wall times spent in each portion of the code
for calculations on the HOMO state of $\textrm{S}\textrm{i}_{\textrm{8381}}\textrm{H}_{\textrm{1620}}$,
with different numbers of Monte Carlo samples. }
\end{figure}

We also performed a set of tests to measure the efficiency of parallelizing
over Monte Carlo samples for the HOMO calculation of the largest $\textrm{S}\textrm{i}_{\textrm{8381}}\textrm{H}_{\textrm{1620}}$
system (Figure \ref{fig:BigTimings}). Here, the number of samples,
$N_{\zeta}$, is set equal to both the number of MPI ranks and the
number of GPUs, so this test is a measure of the weak scaling of the
code. Note that the total runtimes are all $\sim$2500 s, similar
to the full 1024-sample run, demonstrating nearly ideal scaling with
number of samples. 

From Figure \ref{fig:BigTimings}, one can see that the portions of
the code that have been ported to GPUs collectively account for $\sim$38
\% of the total runtime. The remaining, unported portion of the code
includes I/O operations, preparation of the grid representation of
the Hamiltonian, constructing the $\left\{ \zeta\right\} $ and $\left\{ \xi\right\} $
orbitals prior to filtering, solving the linear system to generate
the filter coefficients \cite{gapfilter}, and collecting and post-processing
the samples to produce the final QP energies. 

We also measured the speedup factors for the individual GPU-ported
sections of the code relative to their CPU counterparts. A full-scale
run of the original CPU code is not possible for the larger clusters
due to the wall time limit on NERSC-Perlmutter, so we instead performed
several comparative tests focusing on individual portions of the code
(Table \ref{tab:speedup}). We ran each test for the HOMO state of
$\textrm{S}\textrm{i}_{\textrm{8381}}\textrm{H}_{\textrm{1620}}$
on a single MC sample with other parameters same as in Table \ref{tab:sGWparams}
except that here we decreased the filter length by a factor of 16
(to $N_{chb}=512$) to reduce the CPU time needed for this step. As
Table \ref{tab:speedup} shows, the GPU filtering step achieves a
$\sim50\times$ speedup over its CPU counterpart. The propagation
and spectral estimation steps achieve even higher speedups of $150-250\times$.
This is due not only to porting the routines to GPUs but also to optimizations
that were not present in the CPU code, such as premultiplying potential
energy factors before offloading them to the GPU. The step to generate
the $\left\{ \xi\right\} $ segments, while requiring much less time
than the propagation and filtering steps, exhibited the largest performance
improvement resulting from replacing the serial calls to the KISS
random number generator with calls to the $\mathsf{cuRAND}$ library.
Finally, the last two rows of Table \ref{tab:speedup} list the timings
of the unported code and the sum of all timings, including unported
portions of the code, showing that the overall speedup of the GPU
implementation of $\mathtt{StochasticGW}$ is $\sim45\mathcal{\times}$
that of the CPU code.

\begin{table}
\caption{\label{tab:speedup}Timings and speedups of GPU portions of $\mathtt{StochasticGW}$
relative to the CPU portions, for calculation on the HOMO state of
$\textrm{S}\textrm{i}_{\textrm{8381}}\textrm{H}_{\textrm{1620}}$.}

\begin{tabular}{cccc}
\hline 
Portion & $t_{CPU}$ (s) & $t_{GPU}$ (s) & $\nicefrac{t_{CPU}}{t_{GPU}}$\tabularnewline
\hline 
Propagate $\left\{ \zeta\right\} $ & 4947 & 31 & 160\tabularnewline
Propagate $\left\{ \eta\right\} $ & 37711 & 153 & 246\tabularnewline
Generate $\left\{ \xi\right\} $ & 662 & 0.08 & 8764\tabularnewline
Filter $\left\{ \zeta\right\} $,$\left\{ \eta\right\} $ & 9796 & 199 & 49\tabularnewline
Estimate $\left[E_{min},E_{max}\right]$ & 26364 & 191 & 138\tabularnewline
\hline 
Unported code & \multicolumn{2}{c}{1237} & 1\tabularnewline
Total (incl. unported) & 81292 & 1811 & 45\tabularnewline
\hline 
\end{tabular}
\end{table}

\section{Conclusion}

In this work, we describe a new implementation of the $\mathtt{StochasticGW}$
code. Our code utilizes the stochastic Resolution of the Identity
(sROI) \cite{BreakingLimitsGW,swiftGW} technique, which allows one to decouple the main steps of the
GW method into independent, statistical operations that can be performed
massively in parallel. In deterministic GW methods, the cost is dominated
by computing matrix elements over indices representing the occupied
and unoccupied orbitals. In contrast, in the stochastic method the
cost depends on operations over the full spatial grid and accumulating
a sufficient number of Monte Carlo samples to achieve a desired statistical
accuracy. Therefore, compared to deterministic GW, the cost of stochastic
GW grows much more slowly with respect to the size of the molecule
or material system of interest.

Motivated by the large-scale parallelism available in modern GPU hardware,
we have ported the major computational motifs of the algorithm to
GPUs. These include estimating the spectral width of the Kohn-Sham
Hamiltonian, filtering the initial orbitals by projecting onto an
occupied subspace, and propagating the orbitals under the influence
of a time-dependent Hamiltonian. Each of these steps is applied via
a sequence of vectorized OpenACC kernels and calls to GPU-optimized
FFT libraries.

We showcased the GPU implementation by computing the quasi-particle
energies of the HOMO and LUMO orbitals of five hydrogen-passivated
silicon clusters. The band gaps show convergent behavior towards a
bulk-like limit at ca. 1.36 eV. QP calculations on the largest system,
$\textrm{S}\textrm{i}_{\textrm{8381}}\textrm{H}_{\textrm{1620}}$,
with 10001 atoms and 35144 electrons, can be completed in only $\sim45$
minutes with the workload partitioned with one MC sample per GPU.
For this system, the GPU version of $\mathtt{StochasticGW}$ achieves
roughly $45\times$ speedup in time-to-solution relative to the CPU
version over the entire execution of the code. This work opens the
way for computing QP energies of even larger systems.

\section{Author information}

\paragraph{Corresponding author}

Daniel Neuhauser $-$ Department of Chemistry and Biochemistry, University
of California Los Angeles, Los Angeles, California 90095, United States;
orcid.org/ 0000-0003-3160-386X; Email: dxn@ucla.edu

\paragraph{Authors}

Phillip Thomas $-$ National Energy Research Scientific Computing
Center (NERSC), Lawrence Berkeley National Laboratory, Berkeley, California
94720, United States; orcid.org/0009-0007-3794-0150; Email: pthomas@lbl.gov

Minh Nguyen $-$ Department of Chemistry and Biochemistry, University
of California, Los Angeles, California 90095, United States

Dimitri Bazile $-$ Department of Chemistry and Biochemistry, University
of California Los Angeles, Los Angeles, California 90095, United States

Tucker Allen $-$ Department of Chemistry and Biochemistry, University
of California Los Angeles, Los Angeles, California 90095, United States

Barry Y. Li $-$ Department of Chemistry and Biochemistry, University
of California Los Angeles, Los Angeles, California 90095, United States

Wenfei Li $-$ Department of Chemistry and Biochemistry, University
of California Los Angeles, Los Angeles, California 90095, United States

Mauro Del Ben $-$ Applied Mathematics and Computational Research
Division, Lawrence Berkeley National Laboratory, Berkeley, California
94720, United States; orcid.org/0000-0003-0755-4797

Jack Deslippe $-$ National Energy Research Scientific Computing Center
(NERSC), Lawrence Berkeley National Laboratory, Berkeley, California
94720, United States; orcid.org/0000-0003-1785-4187

\paragraph{Author Contributions}

$^{\S}$ P.T. is first author.

\paragraph{Notes}

The authors declare no competing financial interest.
\begin{acknowledgement}
This work was supported through the Center for Computational Study
of Excited State Phenomena in Energy Materials (C2SEPEM), which is
funded by the U.S. Department of Energy, Office of Science, Basic
Energy Sciences, Materials Sciences and Engineering Division, via
contract no. DE-AC02-05CH11231, as part of the Computational Materials
Sciences Program. Work in Daniel Neuhauser's group was supported by
the National Science Foundation Grant No. CHE2245253. Computational
resources were provided by the National Energy Research Scientific
Computing Center (NERSC), a U.S. Department of Energy Office of Science
User Facility operating under contract no. DE-AC02-05CH11231 using
NERSC awards BES-ERCAP0032245 (project m2651) and DDR-NESAP-ERCAP0033923
(project m4941).
\end{acknowledgement}
\addcontentsline{toc}{section}{\refname}\bibliography{references}

\end{document}